\documentclass[aps,prl,twocolumn,superscriptaddress,showpacs,draft]{revtex4}
\usepackage{graphicx}
\input{epsf}
\usepackage{amssymb}
\usepackage[usenames,dvipsnames]{xcolor}  

\begin{document}

\title{Ecological analysis of world trade}
\author{L.Ermann}
\affiliation{Departamento de F\'isica, 
CNEA, Libertador 8250, (C1429BNP) Buenos Aires, Argentina}
\author{D.L.Shepelyansky}
\affiliation{\mbox{Laboratoire de Physique Th\'eorique du CNRS, IRSAMC, 
Universit\'e de Toulouse, UPS, 31062 Toulouse, France}}

 \date{January 17, 2012 }


%

\maketitle

{\bf Ecological systems have a high level of complexity 
combined with stability and rich biodiversity. Recently, the analysis 
of their properties and evolution has been pushed forward 
on a basis of  concept of mutualistic networks that 
provides a detailed understanding of their features being
linked to a high nestedness of these networks.
It was shown that the nestedness architecture of mutualistic networks of plants 
and their pollinators minimizes  competition and increases
biodiversity. Here, using the United Nations COMTRADE database
for years 1962 - 2009, we show that a similar ecological analysis
gives a valuable description of the world trade.
In fact the countries and trade products are analogous to plants 
and  pollinators, and the whole trade network is characterized
by a low nestedness temperature which is typical for
the ecological networks. This approach provides new mutualistic
features of the world trade 
highlighting new significance of countries and trade products 
for the world trade.
}

Ecological systems are characterized by  high
complexity and biodiversity \cite{maybook}
linked to nonlinear dynamics and chaos 
emerging in the process of their evolution \cite{may1976,ott}.
The interactions between species
form a complex network whose properties can be analyzed 
by the modern methods of scale-free networks
\cite{dorogovtsev,caldarelli1,caldarelli2,dunne}.
An important feature of ecological networks
is that they are highly structured, being very different from
randomly interacting species \cite{dunne,bascompte2003}.
Recently is has been shown that the mutualistic
networks between plants and their pollinators 
\cite{bascompte2003,aizen,bascompte2007pnas,bascompte2007nat}
are characterized by high nestedness  \cite{bascompte2009,bascompte2011}
which minimizes competition and increases biodiversity.
It is argued \cite{bascompte2009} 
that such type of networks appear in 
various social contexts such as
garment industry \cite{bascompte2011} and
banking \cite{maybank,haldanemay}.
Here we apply a nestedness analysis to the world
trade network using the United Nations COMTRADE
database \cite{comtrade} 
for the years 1962 - 2009. Our analysis shows
that  countries and trade products
have relations similar to those of
plants and pollinators and that 
the world trade network is 
characterized by a high nestedness
typical of ecosystems \cite{bascompte2007nat,bascompte2009}. 
This provides new mutualistic characteristics
for the world trade.

\vskip 0.4cm
{\bf Results}

\begin{figure}
\centerline{\epsfxsize=8.2cm\epsffile{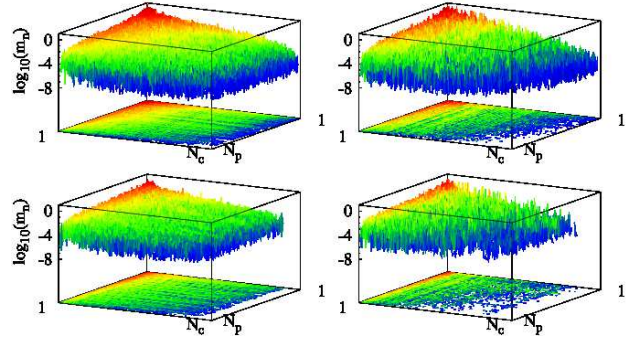}}
\vglue -0.2cm
\caption{Normalized import/export  WTN matrix elements
$m^{(i)}$ and $m^{(e)}$
shown on  left/right panels
for years 1968 (bottom) and 2008 (top).
Here products/countries ($p=1,\ldots,N_p$ and $c=1,\ldots,N_c$)
are ordered in a decreasing order
of product/country total import or export in a given year.
} 
\label{fig1}
\end{figure}

\noindent
The mutualistic World Trade Network (WTN)  is constructed on the
basis of the UN COMTRADE database \cite{comtrade} from the 
matrix of trade transactions $M^p_{c^\prime,c}$ expressed in USD
for a given product
(commodity) $p$ from country $c$ to country $c^\prime$
in a given year (from 1962 to 2009). For product
classification we use  3--digits
Standard International Trade Classification (SITC) Rev.1
with the number of products $N_p=182$.
All these products are described in \cite{comtrade}
in the commodity code document SITC Rev1.
The number of countries varies between $N_c=164$ in 1962
and $N_c=227$ in 2009. The import and export trade matrices
are defined as $M^{(i)}_{p,c}=\sum_{c^\prime=1}^{N_c} M^p_{c,c^\prime}$
and $M^{(e)}_{p,c}=\sum_{c^\prime=1}^{N_c} M^p_{c^\prime,c}$ respectively.
We use the dimensionless matrix elements
$m^{(i)}=M^{(i)}/M_{max}$ and $m^{(e)}=M^{(e)}/M_{max}$ where for a given year
$M_{max}=max\{max[M^{(i)}_{p,c}],max[M^{(e)}_{p,c}]\}$.
The distribution of matrix elements $m^{(i)}$, $m^{(e)}$
in the  plane of indexes $p$ and $c$,
ordered by the total amount of import/export
in a decreasing order, are shown in Fig.~\ref{fig1}
for years 1968 and 2008 (years 1978, 1988, 1998
are shown in Fig.~\ref{figS1} of Supporting Information (SI)). 
These Figs. show that globally the distributions of $m^{(i)}$, $m^{(e)}$
remain stable in time especially in a view of 
100 times growth of the total trade volume 
during the period 1962-2009. The fluctuations of $m^{(e)}$
are visibly larger compared to $m^{(i)}$ case since
certain products, e.g. petroleum, are exported by only a few countries 
while it is imported by almost all countries.

To use the methods of ecological analysis we 
construct the mutualistic network matrix
for import  $Q^{(i)}$ and export $Q^{(e)}$
whose matrix elements take binary value 
$1$ or $0$ if  corresponding elements
 $m^{(i)}$ and $m^{(e)}$ are  respectively larger or smaller
than a certain trade threshold value $\mu$. 
The fraction $\varphi$
of nonzero matrix elements
varies smoothly 
in the range $10^{-6} \leq \mu \leq 10^{-2}$ 
(see Fig.~\ref{figS2} of SI)
and the further analysis is not really sensitive
to the actual $\mu$ value inside this broad range.
It is important to note that in contrast to  ecological
systems \cite{bascompte2009} the world trade is
described by a directed network and hence we 
characterize the system by two 
mutualistic matrices  $Q^{(i)}$ and $Q^{(e)}$ corresponding
to import and export. Using the standard nestedness 
BINMATNEST algorithm \cite{binmatnest} 
we determine the nestedness parameter  $\eta$ of the 
WTN and the related nestedness temperature $T=100 (1-\eta)$.
The algorithm reorders lines and columns of
a mutualistic matrix   
concentrating nonzero elements as 
much as possible in the top left corner
and thus providing information about causal role of
immigration and extinction in an ecological system.
A high level of nestedness and ordering
can be reached only for systems with low $T$.
It is argued that the nested architecture of real mutualistic networks
increases their biodiversity.
\begin{figure}
\centerline{\epsfxsize=8.2cm\epsffile{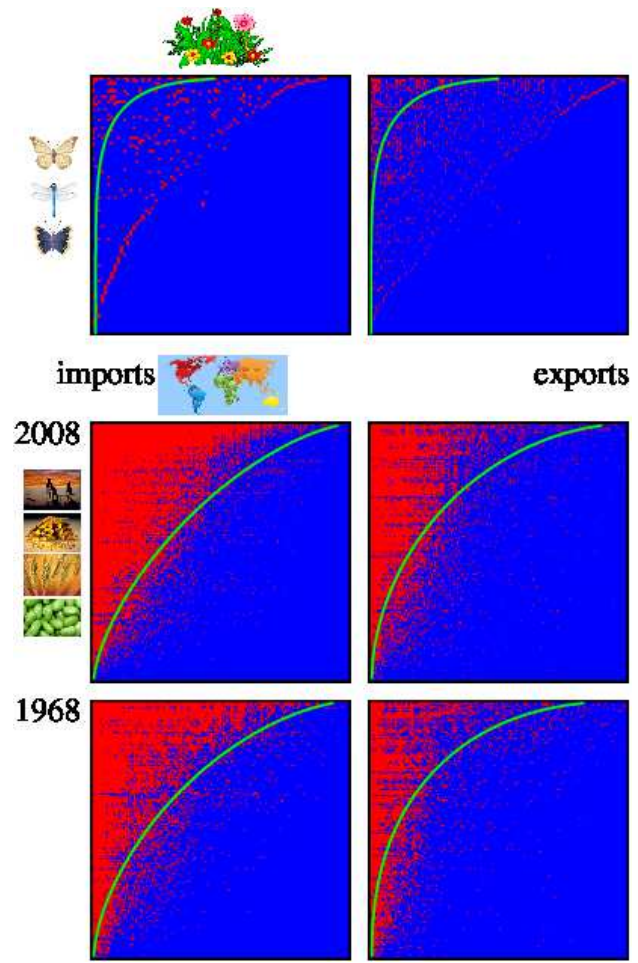}}
\vglue -0.2cm
\caption{
Nestedness matrices for the plant-animal mutualistic 
networks on top panels,  and for the WTN of countries-products 
on middle and bottom panels. Top--left and top--right panels 
represent data of \emph{ARR1} and \emph{WES} networks from
\cite{bascompte2007nat,bascomptedata}.
The WTN matrices are computed with the threshold $\mu=10^{-3}$
and corresponding $\varphi \approx 0.2$
for years 1968 (bottom) and 2008 (middle)  
for import (left panels) and export (right panels).
Red and blue represent unit and zero elements respectively;
only lines and columns with nonzero elements are shown.
The order of plants-animals, countries-products is given by 
the nestedness algorithm \cite{binmatnest}, 
the perfect nestedness is shown 
by green curves for the corresponding values of $\varphi$.
} 
\label{fig2}
\end{figure}

\begin{figure}[h!]
\centerline{\epsfxsize=8.2cm\epsffile{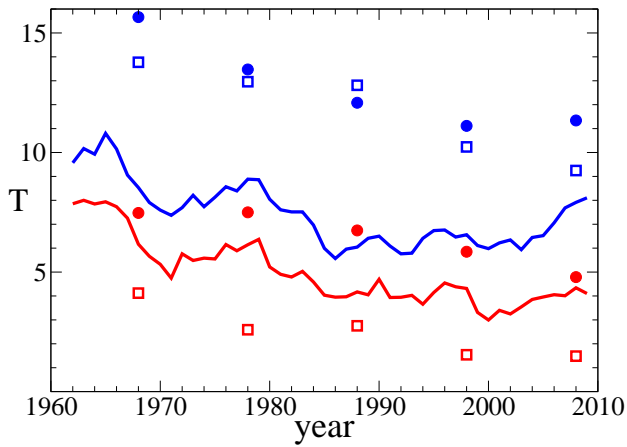}}
\vglue -0.2cm
\caption{Nestedness temperature $T$ as a function of years for the WTN
for $\mu=10^{-3}$ (curves), $10^{-4}$ (circles),
$10^{-6}$ (squares); import and export data are shown in red and blue.
} 
\label{fig3}
\end{figure}

\begin{figure}
 \centerline{\epsfxsize=8.2cm\epsffile{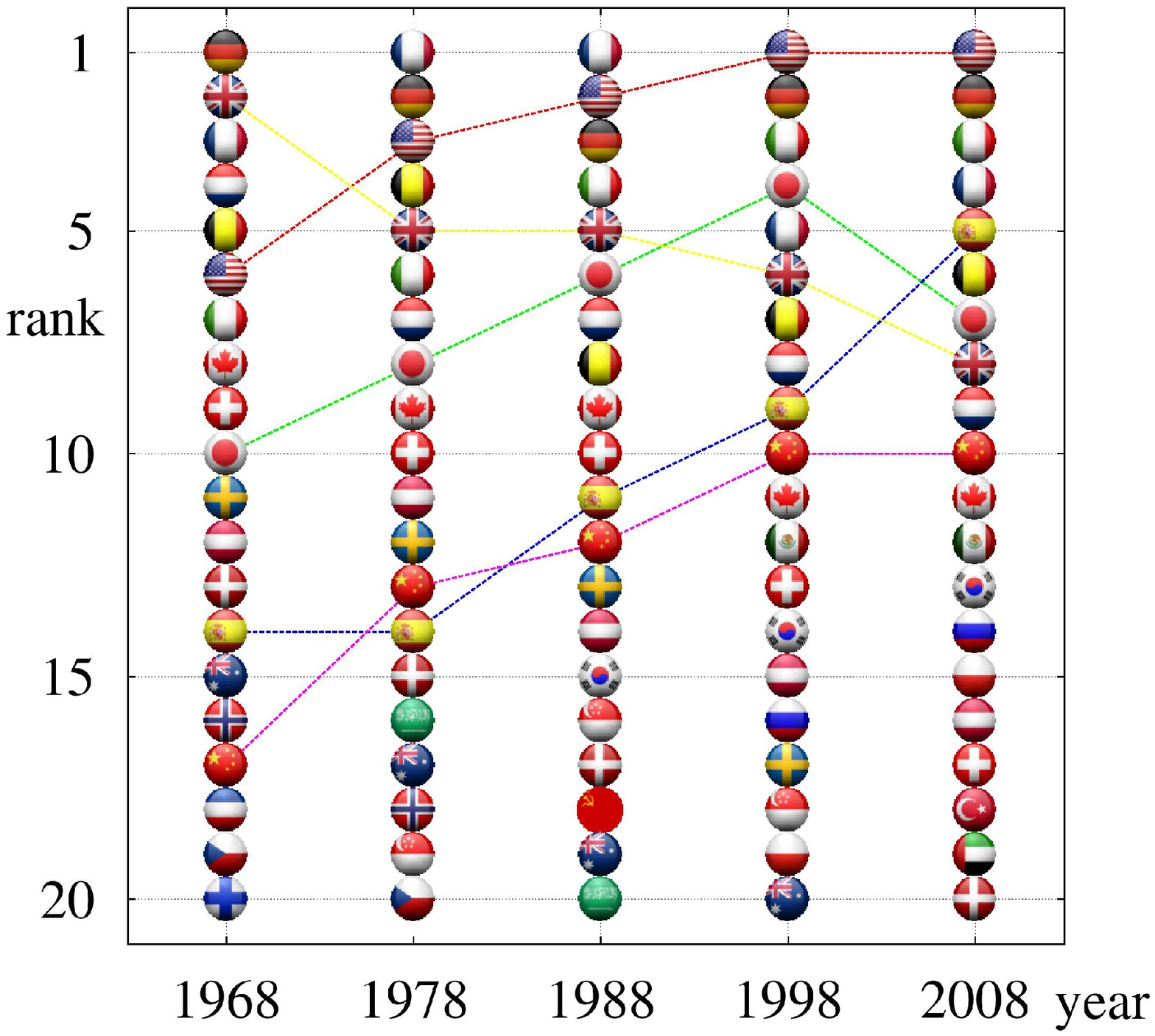}}
 \centerline{\epsfxsize=8.2cm\epsffile{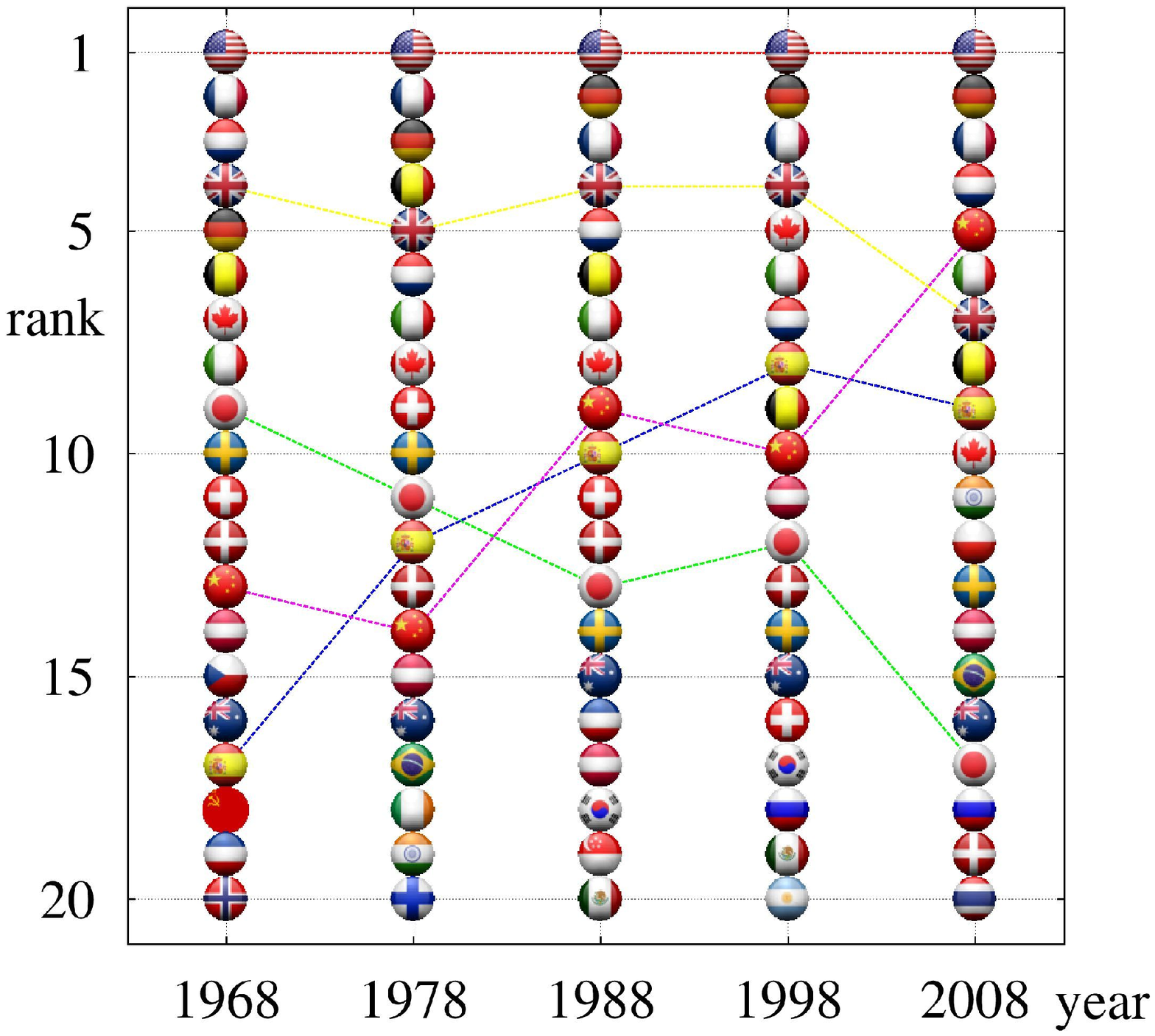}}
 \vglue -0.2cm
 \caption{Top 20 EcoloRank countries as a function of years 
 for the WTN
 import/export on top/bottom 
 panels.   
The ranking is given by the nestedness algorithm \cite{binmatnest}
for the trade threshold $\mu=10^{-3}$; 
each country is represented by its corresponding flag.
As an example, dashed lines show time evolution of  
the following countries:
USA, UK, Japan, China, Spain.
 } 
 \label{fig4}
 \end{figure}

\begin{figure}
 \centerline{\epsfxsize=8.2cm\epsffile{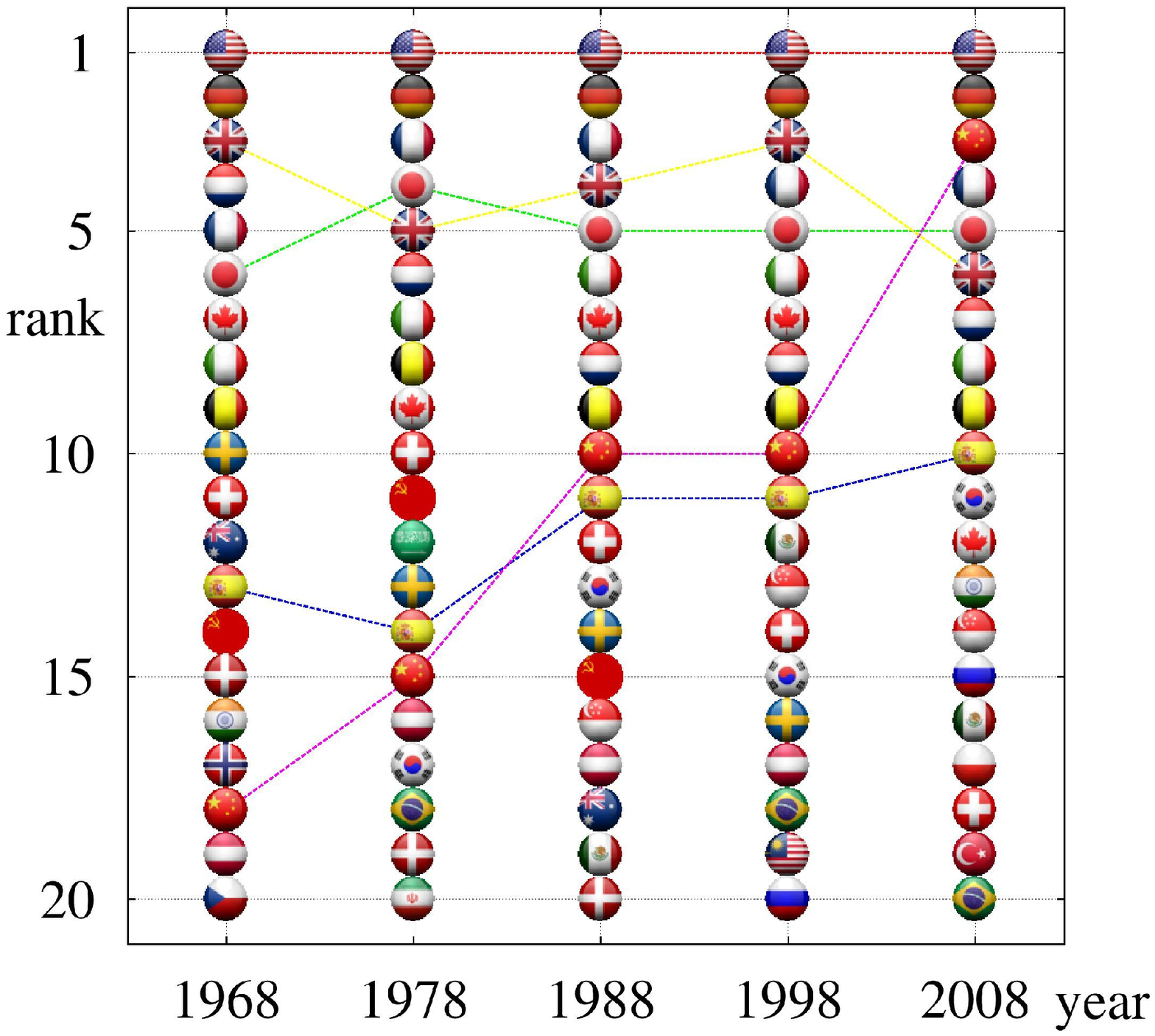}}
 \centerline{\epsfxsize=8.2cm\epsffile{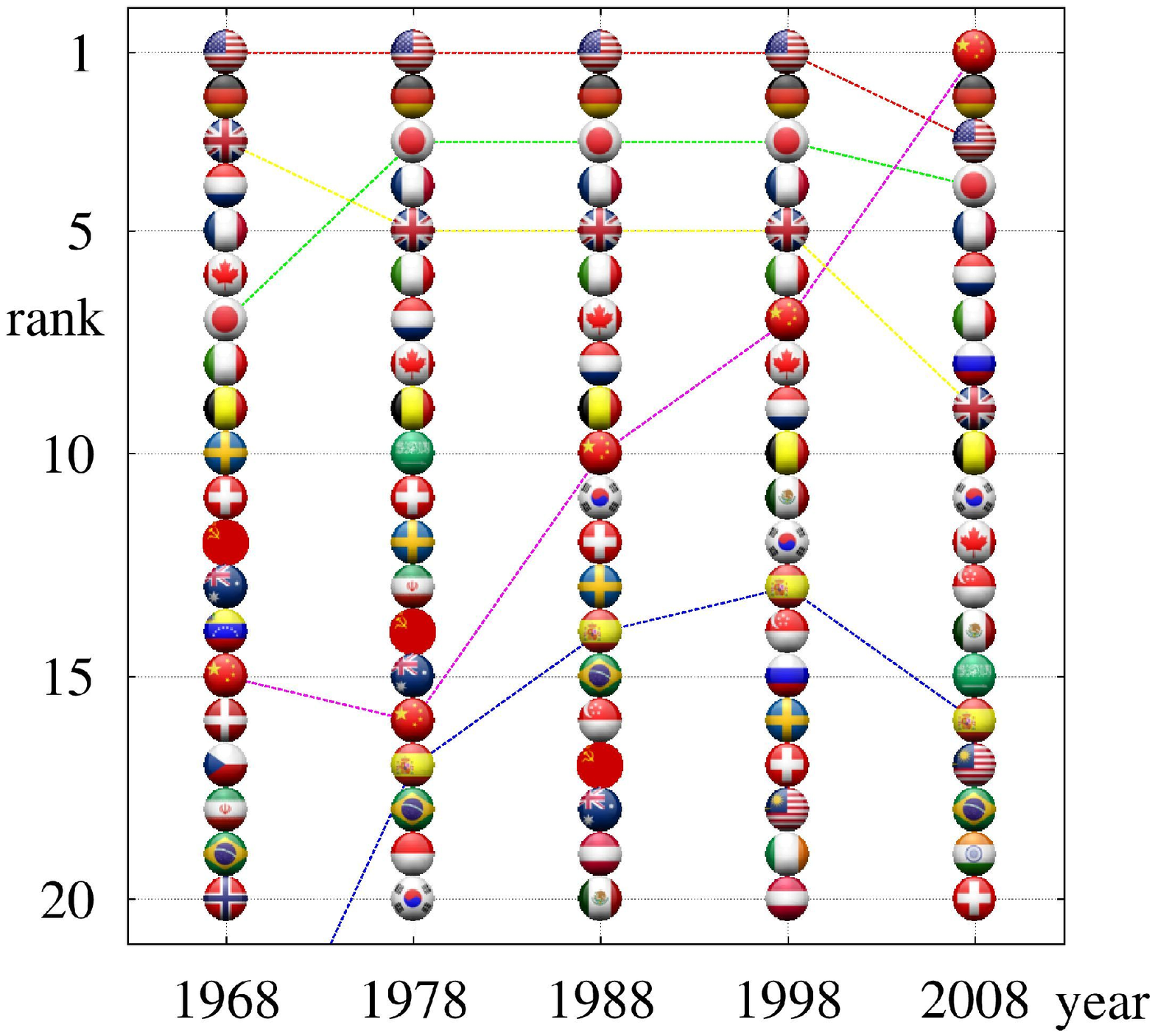}}
 \vglue -0.2cm
 \caption{Top 20 countries as a function of years
ranked by the total monetary trade volume of the WTN
in  import/export  on top/bottom 
 panels respectively; 
each country is represented by its corresponding flag.
Dashed lines show time evolution of  the same countries
as in Fig.~\ref{fig4}.
 } 
 \label{fig5}
 \end{figure}

The nestedness matrices generated by the BINMATNEST algorithm \cite{binmatnest} 
are shown in Fig.~\ref{fig2} for ecology networks
ARR1 ($N_{pl}=84$, $N_{anim}=101$,
$\varphi=0.043$, $T=2.4$) and 
WES ($N_{pl}=207$, $N_{anim}=110$,
$\varphi=0.049$, $T=3.2$) from \cite{bascompte2007nat,bascomptedata}.
Using the same algorithm we generate the nestedness matrices
of WTN using the mutualistic matrices for import 
$Q^{(i)}$ and export $Q^{(i)}$
for the WTN in years 1968 and 2008 using 
a fixed typical threshold  $\mu=10^{-3}$
(see Fig.~\ref{fig2}; the distributions for other $\mu$-values
have a similar form and  are shown
in Fig.~\ref{figS3} of SI). 
As for ecological systems,
for the WTN data we
also obtain rather small nestedness temperature
($T \approx 6/8$ for import/export in 1968 
and $T\approx 4/8$ in 2008 respectively). These values are by
a factor 9/4 of times smaller than the corresponding $T$
values for import/export
from random generated networks with the corresponding 
values of $\varphi$. 
The detailed data for 
$T$ in all years are shown in Fig.~\ref{fig3} 
and the comparison with the 
data for random networks are given in 
Figs.~\ref{figS4},\ref{figS5},\ref{figS6} in SI.
The data of Fig.~\ref{fig3}
show that the value of $T$ changes by about 30-40\%
with variation of $\mu$ by a factor $1000$.
We think that this is relatively small variation
of $T$ compared to enormous variation of $\mu$
that confirms the stability and relevance of
ecological analysis and nestedness ordering.
The nestedness temperature  $T$ remains
rather stable in time: in average
there is 40\% drop of $T$ from 1962 to 2000
and 20\% growth from 2000 to 2009. 
We attribute the growth in last decade to the
globalization of trade.
The small value of nestedness temperature obtained for the WTN
confirms the validity of the ecological analysis
of WTN structure: trade products play the role of pollinators
which produce exchange between world countries, which play the role 
of plants. Like in ecology the WTN
evolves to the state with very low
nestedness temperature that satisfies the
ecological concept
of system stability appearing as a result of high
network nestendess \cite{bascompte2009}.
\begin{figure}
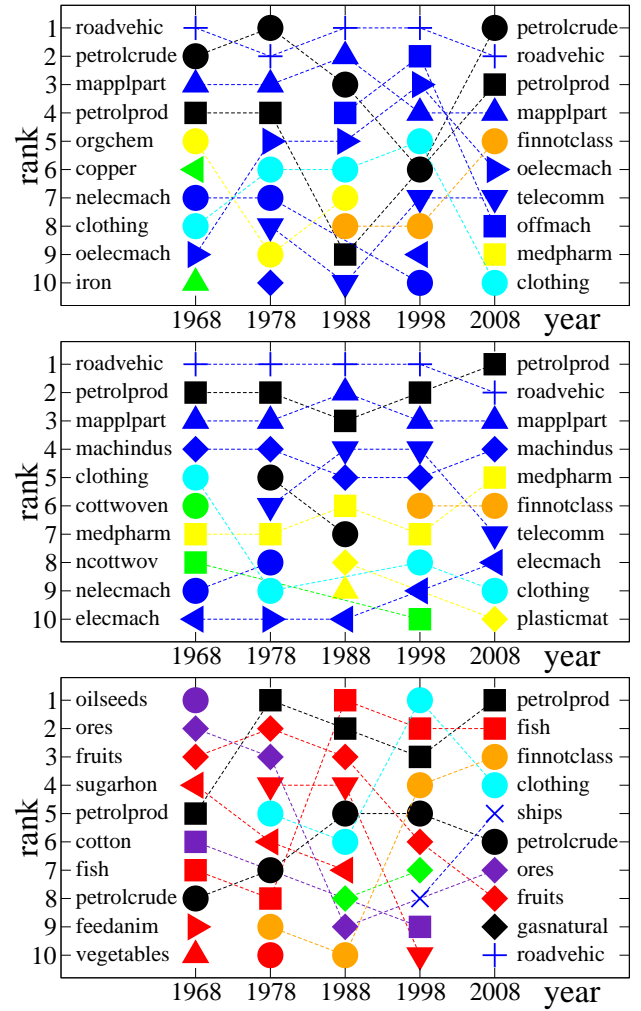

 \centerline{\epsfxsize=8.2cm\epsffile{fig6a.eps}}
 \centerline{\epsfxsize=8.2cm\epsffile{fig6b.eps}}
 \centerline{\epsfxsize=8.2cm\epsffile{fig6c.eps}}
 \vglue -0.2cm
 \caption{Top 10 ranks of trade products as a function of years 
for the WTN. Top panel: ranking of products by
monetary trade volume;
middle/bottom panels:  
ranking is given by the nestedness algorithm \cite{binmatnest}
for import/export
with the trade threshold $\mu=10^{-3}$.
Each product is shown by its own symbol with
short name written at years 1968, 2008;
symbol color marks  1st SITC digit;
SITC codes of products and their names 
are given in Table~\ref{table2} of SI.
 } 
 \label{fig6}
 \end{figure}

The nestedness algorithm \cite{binmatnest}
creates effective ecological ranking  (EcoloRanking) of all UN countries.
The evolution of 20 top ranks throughout the years is shown 
in Fig.~\ref{fig4} for import and export.
This ranking is quite different from the more commonly applied
ranking of countries by their total import/export monetary trade 
volume  \cite{cia2010} (see corresponding data in Fig.~\ref{fig5}) 
or recently proposed 
democratic ranking of WTN based on 
the Google matrix analysis \cite{wtngoogle}.
Indeed, in 2008 China is at the top rank for total export volume
but it is only at 5th position in EcoloRank
(see Fig.~\ref{fig4}, Fig.~\ref{fig5} and  Table I in SI).
In a similar way Japan moves down from 4th to 17th position
while USA raises up from 3rd to 1st rank. 

The same nestedness
algorithm generates not only the ranking of countries
but also the ranking of trade prooducts for 
import and export which is presented in Fig.~\ref{fig6}.
For comparison we also show there the standard ranking of 
products by their trade volume. In  Fig.~\ref{fig6}
the color of symbol marks the 1st SITC digit
described in \cite{comtrade} and in Table~\ref{table2} in SI.

\vskip 0.4cm
{\bf Discussion}

\noindent
The origin of such a difference between EcoloRanking 
and trade volume ranking of countries
is related to the main
idea of mutualistic ranking in ecological systems:
the nestedness ordering stresses the importance of mutualistic 
pollinators (products for WTN)
which generate links and exchange between plants
(countries for WTN). In this way generic products, 
which participate in the trade between many countries,
become of primary importance even if
their trade volume is not at the top lines
of import or export. In fact such mutualistic products
glue the skeleton of the world trade
while the nestedness concept allows to rank them
in order of their importance. The time evolution
of this EcoloRanking of products of WTN
is shown in Fig.~\ref{fig6} for import/export
in comparison with the product ranking by the monetary
trade volume
(since the trade matrix is
diagonal in product index the ranking of 
products in the latter case is the same for
import/export). The top and middle panels
have dominate colors corresponding to
machinery (SITC 7; blue) and mineral fuels (3; black)
with a moderate contribution of chemicals (5; yellow)
and manufactured articles (8; cyan) and a small 
fraction of goods classified by material (6; green).
Even if the global structure of product 
ranking by trade  volume has certain similarities
with import EcoloRanking there are also important 
new elements. Indeed, in 2008 the mutualistic significance of
petroleum products (SITC 332),
{\it machindus} (machines for special industries  718) 
and {\it medpharm}  (medical-pharmaceutic products  541) 
is much higher
compared to their volume ranking, while 
petroleum crude (331) and 
office machines (714) have smaller mutualistic significance
compared to their volume ranking.

The new element of EcoloRanking is that it differentiates 
between import and export products while for trade volume
they are ranked in the same way. Indeed, the dominant colors
for export (Fig.~\ref{fig6}  bottom panel)
correspond to food  (SITC 0; red) with contribution 
of black (present in import) and crude materials (2; violet);
followed by cyan (present in import) and more pronounced presence of
{\it finnotclass} (commodities/transactions not classified 9; brown).
EcoloRanking of export shows a clear decrease tendency
of dominance of SITC0 and SITC2 with time and increase
of importance of SITC3,7. It is interesting to note that
petroleum products SITC332 is vary vulnerable
in volume ranking
due to significant variations of petroleum prices  
but in EcoloRanking this product keeps the stable
top positions in all years showing its 
mutualistic structural importance for the world trade.
EoloRanking of export shows also importance of
fish (SITC031), clothing (SITC841) and fruits (SITC051)
which are placed on higher positions compared to their volume ranking.
At the same time {\it roadvehic} (SITC732),
which are at top volume ranking, have 
relatively low ranking in export since only a few countries
dominate the production of road vehicules.

It is interesting to note that in Fig.~\ref{fig6}
petroleum crude is at the top of trade volume ranking e.g. in 2008 (top panel)
but it is absent in import EcoloRanking (middle panel)
and it is only on 6th position in export EcoloRanking (bottom panel).
A similar feature is visible for years 1968, 1978.
On a first glance this looks surprising but in fact for 
mutualistic EcoloRanking it is important that 
a given product is imported from top EcoloRank countries:
this is definitely not the case for petroleum crude
which practically is not produced inside
top 10 import EcoloRank countries (the only exception is USA,
which however also does not export much). Due to that reason
this product has low mutualistic significance.

The mutualistic concept of product importance 
is at the origin of significant difference
of EcoloRanking of countries compared to the usual trade
volume ranking 
(see Fig.~\ref{fig4}, Fig.~\ref{fig5}).
Indeed, in the latter case China and Japan are at the dominant
positions but their trade is concentrated in
specific products which mutualistic role is 
relatively low. In contrast USA, Germany and France
keep top three EcoloRank positions 
during almost 40 years clearly demonstrating their mutualistic
power and importance for the world trade.

In conclusion, our results show the universal features of ecologic
ranking of complex networks with promising future applications
to trade, finance and other areas.

{\bf Acknowledgments:} We thank Arlene Adriano and Matthias Reister
(UN COMTRADE) for  provided help and friendly access to the
database \cite{comtrade}.


\renewcommand{\theequation}{A-\arabic{equation}}
  \setcounter{equation}{0}  
\renewcommand{\thefigure}{S-\arabic{figure}}
  \setcounter{figure}{0}  

\section{Supporting  Information}
Here we present the Supporting Information (SI) 
for the main part of the paper.

\begin{figure}[h!]
\centerline{\epsfxsize=8.2cm\epsffile{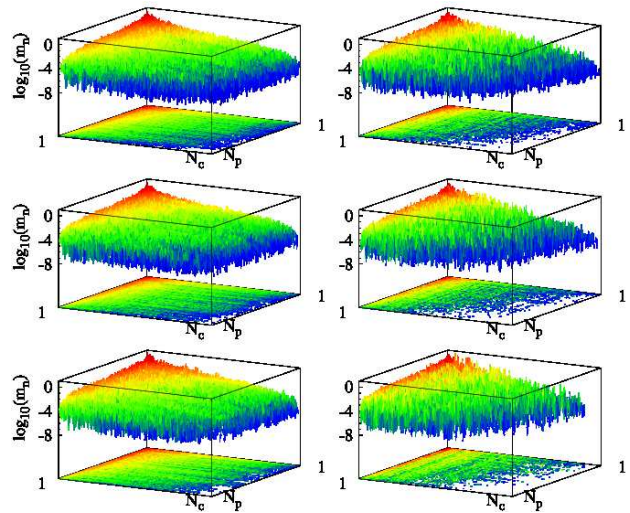}}
\vglue -0.2cm
\caption{Same type of WTN matrix data as in Fig.~\ref{fig1}
shown for years  1978, 1988, 1998 
in panels from bottom to top respectively.
} 
\label{figS1}
\end{figure}

In Fig.~\ref{figS1}, in a complement to Fig.~\ref{fig1},
we show the normalized WTN matrix
for import $m^{(i)}$ and export $m^{(e)}$
at additional year 1978,1988,1998. As in Fig.~\ref{fig1}
all products and countries are ordered in a decreasing order
of product $(p=1, \ldots, N-p)$ and country
$(c=1,\ldots,N_c)$ import (left panels) and export (right panels)
in a given year. These data show that the global distribution 
remains stable in time: indeed, the global monetary trade volume
was increased by a factor 100 from year 1962 to 2008 
(see e.g. Fig.~5 in \cite{wtngoogle}) but the shape of the distribution
remained essentially the same.

\begin{figure}[h!]
\centerline{\epsfxsize=8.2cm\epsffile{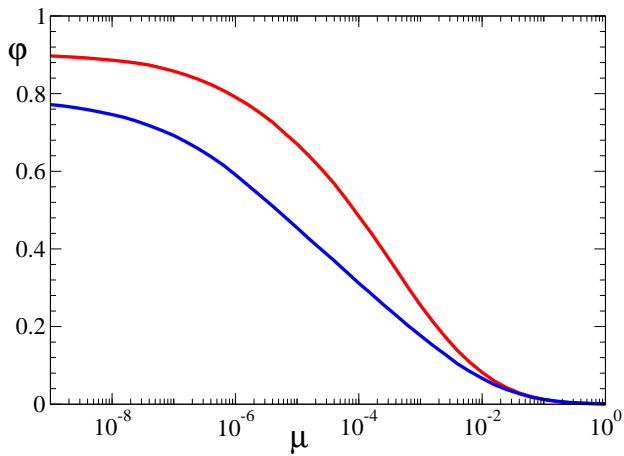}}
\vglue -0.2cm
\caption{The fraction $\varphi$ of nonzero matrix elements 
for the mutualistic network 
matrices of import  $Q^{(i)}$ and  export $Q^{(e)}$
as a function of the cutoff trade threshold $\mu$ for the
normalized WTN matrices $m^{(i)}$ and $m^{(e)}$ for the year 2008; 
the red curve shows the case of import while 
the blue curve shows the case of export network. 
} 
\label{figS2}
\end{figure}

The dependence of the fraction $\varphi$ of nonzero elements
of the mutualistic matrices of import $Q^{(i)}$
and export $Q^{(e)}$ on the cutoff threshold $\mu$
is shown in Fig.~\ref{figS2}. In the range 
of $10^{-6} \leq \mu \leq 10^{-2}$ there is a smooth relatively
weak variation of $\varphi$ with $\mu$.

\begin{figure}[h!]
\centerline{\epsfxsize=8.2cm\epsffile{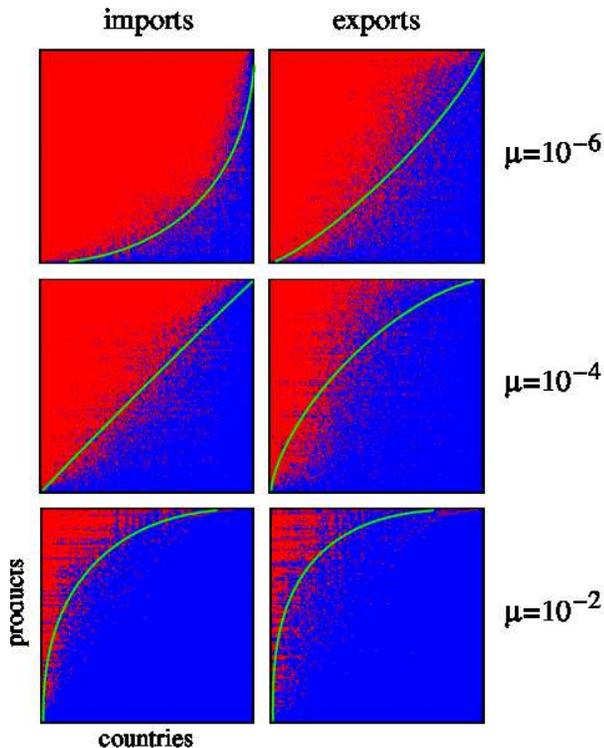}}
\vglue -0.2cm
\caption{Same as in Fig.~\ref{fig2}: 
nestedness matrix for the  WTN data in 2008 
shown for the threshold values $\mu=10^{-6},10^{-4},10^{-2}$)
(from top to bottom);  the perfectly nestedness
is shown by green curves
for the corresponding values of  $\varphi$
taken from Fig.~\ref{figS2}.
} 
\label{figS3}
\end{figure}

In Fig.~\ref{figS3}, in addition to Fig.~\ref{fig2}, 
we show the nestedness matrices of WTN at various values of 
the cutoff threshold $\mu$. The data at various $\mu$ 
values show that in all cases the nestedness algorithm
\cite{binmatnest} correctly generates a matrix 
with nestedness structure.


The variation of the nestedness temperature $T$
with time is shown in Fig.~\ref{fig3}
and several values of the trade threshold $\mu$.
These data show that in average the value of $T$
for export is higher than for import.
We attribute this to stronger fluctuations of
matrix elements of $m^{(e)}$ compared
to those of   $m^{(e)}$ that is well visible in
Figs.~\ref{fig1},\ref{figS1}. As it is pointed
in the main part, we attribute this to the fact that e.g.
only a few countries export petroleum crude
while the great majority of countries import
this product. 

\begin{figure}[h!]
\centerline{\epsfxsize=8.2cm\epsffile{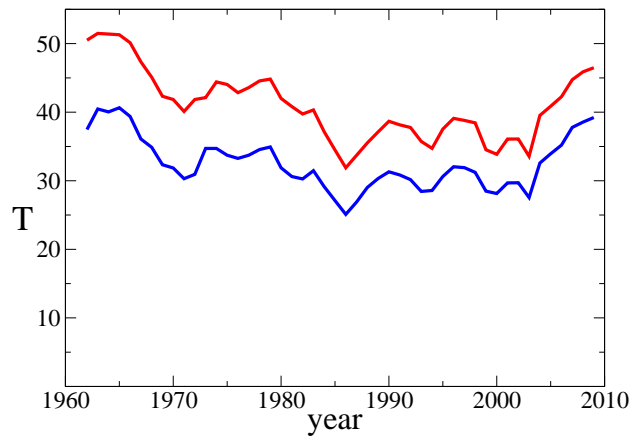}}
\vglue -0.2cm
\caption{Nestedness temperature $T$ for the model given by 
random generated networks; 
here $T$ is computed with 500 random realisations of network
for each year 
using $N_p$, $N_c$ and $\varphi$ of the corresponding WTN data
in this year at $\mu=10^{-3}$; 
import/export data are shown by red/blue curves respectively.
} 
\label{figS4}
\end{figure}

\begin{figure}
\centerline{\epsfxsize=8.2cm\epsffile{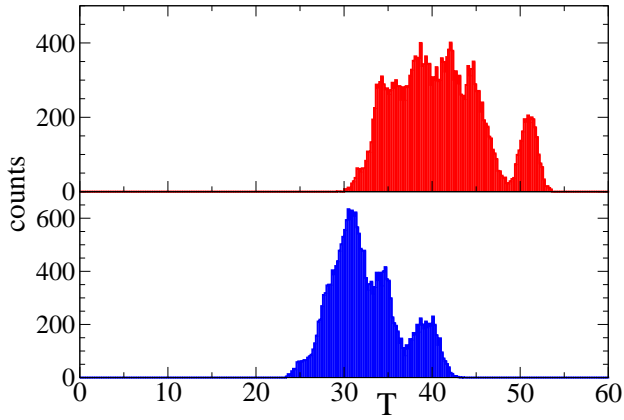}}
\vglue -0.2cm
\caption{Histogram of temperatures for 500 random generated networks
per year (from 1962 to 2009). Top (bottom) panel represents import (export) 
data; here the
parameter values of $N_p$, $N_c$ and $\varphi$ are as for 
the corresponding WTN years at $\mu=10^{-3}$.
} 
\label{figS5}
\end{figure}

In Fig.~\ref{figS4} we show the nestedness temperature 
dependence on time
for the case of random generated networks which have the same 
fraction of nonzero matrix elements $\varphi$
as the WTN at the given year and $\mu=10^{-3}$.
These data, compared with those of Fig.~\ref{fig3}, 
really demonstrate that
the real WTN has values of $T$ by a factor 5 (export)
to 10 (import) smaller comparing to the random networks.
This confirm the nestedness structure of WTN being
similar to the case of ecology networks discussed in
\cite{bascompte2009}. It is interesting to note that
for random generated networks the values of $T$
for  import are larger than for export
while to the WTN we have the opposite relation.
The histogram of distribution of $T$ 
for random generated networks for all years
1962-2009 is shown in Fig.~\ref{figS5}.
Even minimal values of $T$ remain several times larger
than the WTN values of $T$. 

\begin{figure}
\centerline{\epsfxsize=8.2cm\epsffile{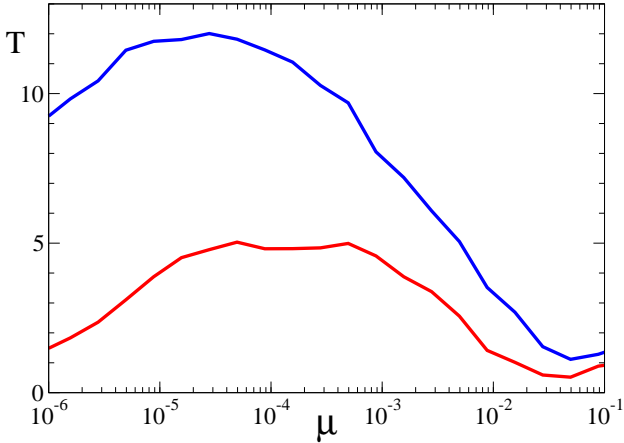}}
\vglue -0.2cm
\caption{Nestedness temperature in the WTN at year 2008 
as a function of threshold 
$\mu$; imports/exports networks are shown  by red/blue curves respectively.
} 
\label{figS6}
\end{figure}

In Fig.~\ref{figS6} we show the dependence of $T$ on 
the trade threshold $\mu$ for the WTN data in year 2008.
We see that there is only about 10-20\% of variation of
$T$ for the range  $10^{-5} \leq \mu  \leq 10^{-3}$.
Even for a much larger range $10^{-6} \leq \mu  \leq 10^{-2}$
the variation of $T$ remains smooth and remains in the
bounds of 100\%. This confirms the stability of
nestedness temperature in respect to broad range variations of $\mu$.
We present the majority of our data for $\mu=10^{-3}$
which is approximately located in the flat range of $T$
variation in year 2008. The data of Table~\ref{table1}
for EcoloRanking of countries at two different values of $\mu$
in year 2008 confirm the stability of this nestedness ordering.  
At the same time larger values of $\mu$ stress the importance of
countries with a large trade volume, e.g. the position of China
in export goes up from rank 5 at $\mu=10^{-3}$ to rank 3 at $\mu=10^{-2}$.

In Table~\ref{table1} we present trade volume ranking
and EcoloRanking of top 20 countries for import/export
of WTN in year 2008.

In Table~\ref{table2} we give the notations and symbols for Fig.~\ref{fig6}
with corresponding SITC Rev1 codes and names. The list
of all SITC  Rev1 codes is available at \cite{comtrade}
(see file 
{\footnotesize \verb|http://unstats.un.org/unsd/tradekb/Attachment193.aspx|}).
The colors of symbols in Fig.~\ref{fig4} mark the first
digit of SITC Rev1 code: 0 - red (Food and live animals);
1 - does not appear in Fig.~\ref{fig4} (Beverages and tobacco);
2 - violet (Crude materials, inedible, except fuels);
3 - black (Mineral fuels, lubricants and related materials);
4 - does not appear in Fig.~\ref{fig4} (Animal and vegetable oils and fats);
5 - yellow (Chemicals);
6 - green (Manufact goods classified chiefly by material);
7 - blue (Machinery and transport equipment);
8 - cyan (Miscellaneous manufactured articles);
9 - brown (Commod. and transacts. Not class. Accord. To kind).

\begin{table*}[h]%
\caption {Top 20 ranks of countries for import and export  
with ranking by the  monetary  trade volume and 
by the nestedness algorithm at two threshold values 
$\mu$ (year 2008).}
\label {table1}\centering %
\begin{tabular}{r*{7}{c}} 
\hline
&  
\multicolumn{3}{c}{import} & 
\multicolumn{3}{c}{export} \\
Rank& Money& $\mu=10^{-3}$ & $\mu=10^{-2}$ &
Money&$\mu=10^{-3}$ &$\mu=10^{-2}$ \\
\hline

1 &USA &USA &USA &China &USA &USA\\
2 &Germany &Germany &Germany &Germany &Germany &Germany\\
3 &China &Italy &France &USA &France &China\\
4 &France &France &UK &Japan &Netherlands &France\\
5 &Japan &Spain &Italy &France &China &Italy\\
6 &UK &Belgium &Netherlands &Netherlands &Italy &Netherlands\\
7 &Netherlands &Japan &Belgium &Italy &UK &Belgium\\
8 &Italy &UK &Japan &Russian Fed. &Belgium &UK\\
9 &Belgium &Netherlands &China &UK &Spain &Japan\\
10 &Canada &China &Spain &Belgium &Canada &Spain\\
11 &Spain &Canada &Canada &Canada &India &Canada\\
12 &Rep. of Korea &Mexico &Russian Fed. &Rep. of Korea &Poland &Switzerland\\
13 &Russian Fed. &Rep. of Korea &Rep. of Korea &Mexico &Sweden &India\\
14 &Mexico &Russian Fed. &Switzerland &Saudi Arab &Austria &Rep. of Korea\\
15 &Singapore &Poland &Austria &Singapore &Brazil &Poland\\
16 &India &Austria &Poland &Spain &Australia &Turkey\\
17 &Poland &Switzerland &Sweden &Malaysia &Japan &Czech Rep.\\
18 &Switzerland &Turkey &Mexico &Brazil &Russian Fed. &Austria\\
19 &Turkey &U. Arab Emir. &India &India &Denmark &Thailand\\
20 &Brazil &Denmark &Singapore &Switzerland &Thailand &Denmark\\
\hline
\end{tabular}
\end{table*}

\begin{table*}[h]%
\caption {Product names for SITC Rev1 3--digit code used in Fig.~\ref{fig4}}
\label {table2}\centering %
\begin{tabular}{cccl}
\hline
Symbol&Code&Abbreviation&Name\\
\hline
{\color{red} \large{$\bullet$}}&001 &animals&Live animals\\
{\color{red} $\blacksquare$}&031 &fish&Fish, fresh \&
simply preserved\\
{\color{red} $\blacklozenge$}&051 &fruits&Fruit, fresh,
and nuts  excl. Oil nuts\\
{\color{red} $\blacktriangle$}&054 &vegetables&Vegetables,
roots \& tubers, fresh or dried\\
{\color{red} $\blacktriangleleft$}&061 &sugarhon&Sugar and
honey\\
{\color{red} $\blacktriangledown$}&071 &coffee&Coffee\\
{\color{red} $\blacktriangleright$}&081 &feedanim&Feed.
stuff for animals excl. unmilled cereals\\
{\color{violet} \large{$\bullet$}}&221 &oilseeds&Oil seeds,
oil nuts and oil kernels\\
{\color{violet} $\blacksquare$}&263 &cotton&Cotton\\
{\color{violet} $\blacklozenge$}&283 &ores&Ores \& concentrates
of non ferrous base metals\\
{\color{black} \large{$\bullet$}}&331 &petrolcrude&Petroleum,
crude and partly refined\\
{\color{black} $\blacksquare$}&332 &petrolprod&Petroleum products\\
{\color{black} $\blacklozenge$}&341 &gas&Gas, natural and
manufactured\\
{\color{yellow} \large{$\bullet$}}&512 &orgchem&Organic chemicals\\
{\color{yellow} $\blacksquare$}&541 &medpharm&Medicinal
\& pharmaceutical products\\
{\color{yellow} $\blacklozenge$}&581 &plasticmat&Plastic
materials,regenerd. cellulose \& resins\\
{\color{yellow} $\blacktriangle$}&599 &chemmat&Chemical
materials and products, nes\\
{\color{green} \large{$\bullet$}}&652 &cottwoven&Cotton
fabrics, woven ex. narrow or spec.fabrics\\
{\color{green} $\blacksquare$}&653 &ncottwov&Text fabrics
woven ex narrow, spec, not cotton\\
{\color{green} $\blacklozenge$}&667 &pearlsprec&Pearls and
precious and semi precious stones\\
{\color{green} $\blacktriangle$}&674 &iron&Universals, plates
and sheets of iron or steel\\
{\color{green} $\blacktriangleleft$}&682 &copper&Copper\\
{\color{blue} \large{$\bullet$}}&711 &nelecmach&Power
generating machinery, other than electric\\
{\color{blue} $\blacksquare$}&714 &offmach&Office machines\\
{\color{blue} $\blacklozenge$}&718 &machindus&Machines for
special industries\\
{\color{blue} $\blacktriangle$}&719 &mapplpart&Machinery and
appliances non electrical parts\\
{\color{blue} $\blacktriangleleft$}&722 &elecmach&Electric
power machinery and switchgear\\
{\color{blue} $\blacktriangledown$}&724 &telecomm&
Telecommunications apparatus\\
{\color{blue} $\blacktriangleright$}&729 &oelecmach& Other
electrical machinery and apparatus\\
{\color{blue} $+$}&732 &roadvehicles&Road motor vehicles\\
{\color{blue} $\times$}&735 &ships&Ships and boats\\
{\color{cyan} \large{$\bullet$}}&841 &clothing&Clothing
except fur clothing\\
{\color{orange} \large{$\bullet$}}&931 &finnotclass& Special
transactions not classd. accord.to kind\\
\hline
\end{tabular}
\end{table*}

\end{document}